\documentclass[aps,prx,twocolumn,superscriptaddress,groupedaddress]{revtex4}  
\usepackage{graphicx}  
\usepackage{dcolumn}   
\usepackage{bm}        
\usepackage{amssymb}   
\usepackage{amsmath}   
\usepackage{color}
\usepackage{ulem}
\begin{document}

\title{Dissipation-driven selection in non-equilibrium chemical networks}
\author{D. M. Busiello}
\affiliation{Institute of Physics, Ecole Polytechnique F\'ed\'erale de Lausanne (EPFL), 1015 Lausanne, Switzerland}
\author{S.-L. Liang}
\affiliation{Institute of Physics, Ecole Polytechnique F\'ed\'erale de Lausanne (EPFL), 1015 Lausanne, Switzerland}
\author{P. De Los Rios}
\affiliation{Institute of Physics, Ecole Polytechnique F\'ed\'erale de Lausanne (EPFL), 1015 Lausanne, Switzerland}
\affiliation{Institute of Bioengineering, Ecole Polytechnique F\'ed\'erale de Lausanne (EPFL), 1015 Lausanne, Switzerland}

\date{\today}

\begin{abstract}
Life has most likely originated as a consequence of processes taking place in non-equilibrium conditions (\textit{e.g.} in the proximity of deep-sea thermal vents) selecting states of matter that would have been otherwise unfavorable at equilibrium. Here we present a simple chemical network in which the selection of states is driven by the thermodynamic necessity of dissipating heat as rapidly as possible in the presence of a thermal gradient: states participating to faster reactions contribute the most to the dissipation rate, and are the most populated ones in non-equilibrium steady-state conditions. Building upon these results, we show that, as the complexity of the chemical network increases, the \textit{velocity} of the reaction path leading to a given state determines its selection, giving rise to non-trivial localization phenomena in state space. A byproduct of our studies is that, in the presence of a temperature gradient, thermophoresis-like behavior inevitably appears depending on the transport properties of each individual state, thus hinting at a possible microscopic explanation of this intriguing yet still not fully understood  phenomenon.
\end{abstract}

\maketitle

\section{Introduction}

The emergence of cellular life has likely been preceded by the appearance of molecular ``replicators", namely molecules able to use basic building blocks present in the environment to create copies of themselves.
RNA and other  long macromolecules, such as proteins, are considered as the best candidates for the first replicators.

Although in the present oxidative conditions long biomolecules such as RNA are not thermodynamically stable, possible more favorable conditions in primordial Earth might have been more conductive to the abiotic synthesis of amino-acids and nucleic acids (see \cite{trailOxidationStateHadean2011,lyonsRiseOxygenEarth2014} for the still-open debate). Nonetheless, no conditions have been found to date such that either the final products or their precursors could be stable and abundant enough to further proceed to their spontaneous polymerization \cite{kimSynthesisCarbohydratesMineralGuided2011} and subsequent self-replication. Relying 
on equilibrium thermodynamics is thus unlikely to provide a route to explain the emergence of life, and possibly it would raise an even more daunting issue: when, precisely, the switch from equilibrium to non-equilibrium replicators, as observed in present life, would have taken place.

A different scenario is the possibility that, from the onset, external sources of energy might have driven pre-biotic molecules away from equilibrium, allowing higher-energy states (\textit{i.e.} more complex and/or longer molecules) to be abundant against their natural tendency to decay according to their equilibrium fate. Consistently with these arguments, Braun and coworkers \cite{mastEscalationPolymerizationThermal2013} have for example shown that, in the presence of thermal gradients, the accumulation of molecules in regions of lower temperature (thermophoresis) increases polymerization beyond the prescriptions of mass-action kinetics at equilibrium.

In the present work we want to broaden the perspective by showing that external energy sources, here a thermal gradient, can tilt the populations of the different states that participate to a reaction network, by favoring the states that take part to faster reactions. In particular we use simple reaction networks to highlight the basic rules deciding which states are the most favourable, relating them to kinetic and dissipation rates. 

\section{Results}

\subsection{A temperature gradient favors states involved in faster reaction pathways.}

\begin{figure}[t]
\centering
\includegraphics[width=\columnwidth]{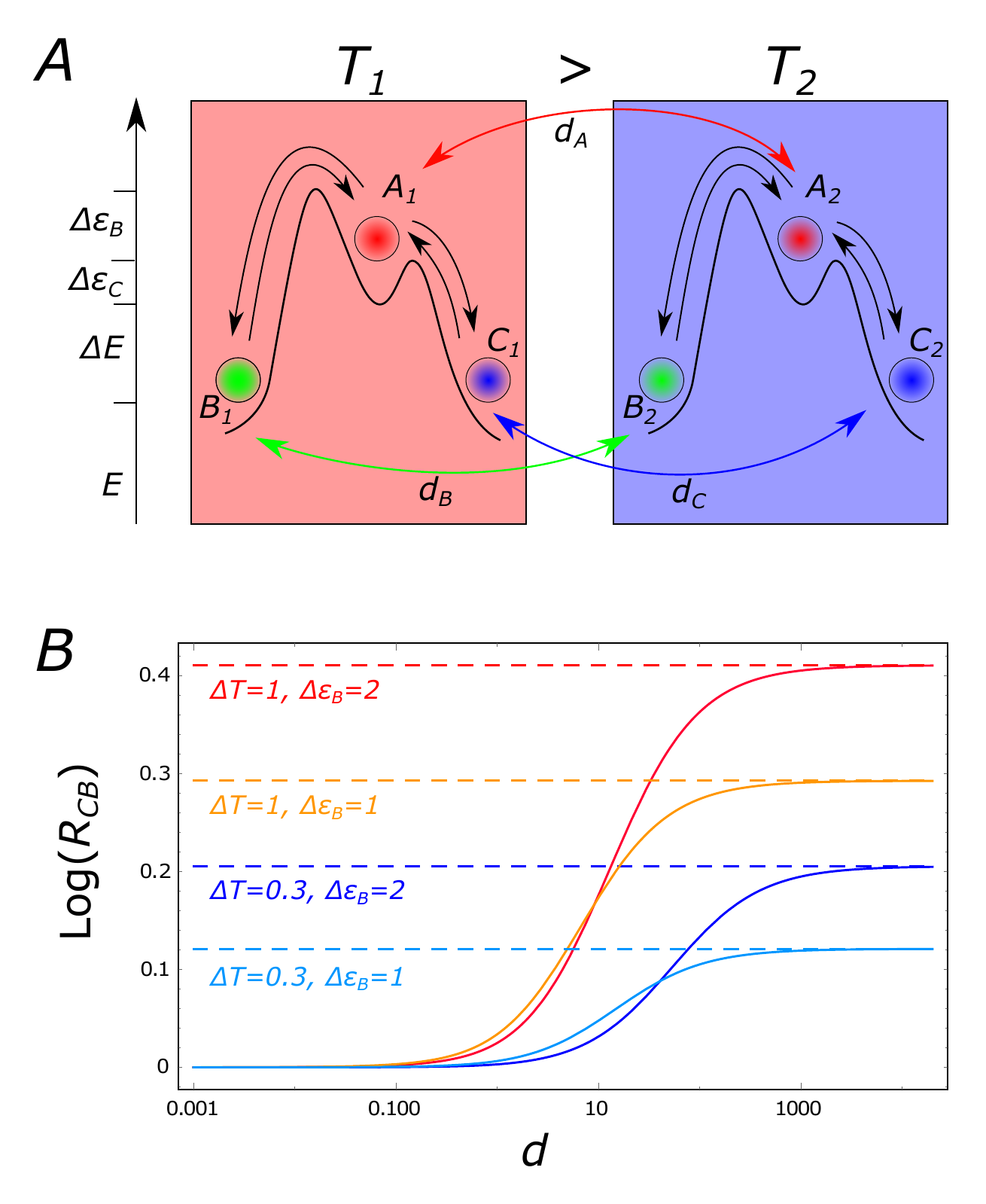}
\caption{A) A three-state chemical system diffusing in a temperature gradient, modeled as two connected  boxes at different temperatures, $T_1 $ and $T_2$ ($T_1-T_2 = \Delta T > 0$). The states $B$ and $C$ have the same energy and the energy barrier between $A$ and $C$, $\Delta \epsilon_C$, is lower than the one between $A$ and $B$, $\Delta \epsilon_B$. B) The quantity $R_{CB} = (P(C_1) + P(C_2))/(P(B_1) + P(B_2))$ gauges the global non-equilibrium unbalance between $B$ and $C$. Under non-equilibrium conditions, $C$ is favored with respect to $B$, since it participates in faster reactions. Here, $T_2=0.5, \Delta E=1, \Delta \epsilon_C = 0, T_1 = T_2 + \Delta T$ and $d=d_A=d_B=d_C$, with $k_B$ set to $1$. $R_{CB}$ is a monotonously increasing function of $d$, and the dashed lines indicate the limit for $d \to \infty$, computed in Eq.~\eqref{RCB}.}
\label{fig:1}
\end{figure}

The simple toy model that we propose here comprises three states, $A$, $B$ and $C$, which diffuse in space in the presence of a temperature gradient $\Delta T$. A pedagogical way to describe this system retaining all its essential non-equilibrium features, is by means of a two-box model as depicted in Fig. \ref{fig:1}A. Here diffusion is captured by allowing each state to move back and forth between the two boxes, with transport rates $d_A$, $d_B$ and $d_C$. The system evolves according to a Master Equation \cite{gardiner, schnakenberg}: 
\begin{eqnarray}
\frac{dP(X_1)}{dt} &=& \sum_{Y_1} \left( k_{Y_1 \to X_1} P(Y_1) - k_{X_1 \to Y_1} P(X_1) \right) + \nonumber \\
&\;& + ~d_X (P(X_{2}) - P(X_1)) \nonumber \\
\frac{dP(X_2)}{dt} &=& \sum_{Y_2} \left( k_{Y_2 \to X_2} P(Y_2) - k_{X_2 \to Y_2} P(X_2) \right) + \nonumber \\
&\;& + ~d_X (P(X_{1}) - P(X_2))
\label{ME}
\end{eqnarray}
where $X,Y=A,B,C$. To take into account the energy differences between the different states, the following relations between the transition rates must be respected  \cite{raz, jarz, astum, bamos1}:
\begin{eqnarray}
k_{A_1 \to B_1} &=& e^{(E_A - E_B)/k_B T_1} k_{B_1 \to A_1} \nonumber \\
k_{A_1 \to C_1} &=& e^{(E_A - E_C)/k_B T_1} k_{C_1 \to A_1} \nonumber \\
k_{A_2 \to B_2} &=& e^{(E_A - E_B)/k_B T_2} k_{B_2 \to A_2} \nonumber \\
k_{A_2 \to B_2} &=& e^{(E_A - E_C)/k_B T_2} k_{B_2 \to A_2}\quad .
\label{arrhenius}
\end{eqnarray}
with $T_1 - T_2 = \Delta T > 0$. To further emphasize the effects that we want to highlight, we set the energies of the states $B$ and $C$ to be equal, $E_B = E_C$ (and $\Delta E = E_A- E_B = E_A - E_C$),
with the additional condition on the height of the barrier that, \textit{\`a la} Arrhenius, determines the velocity of the reactions
\begin{eqnarray}
k_{C_i \to A_i} = e^{\Delta \epsilon /k_B T_i} k_{B_i \to A_i} \qquad \textrm{for } i=1,2
\label{barriers}
\end{eqnarray}
with $\Delta \epsilon = \Delta \epsilon_B -\Delta \epsilon_C > 0$. Eqs.~\eqref{barriers} imply that, irrespective of the temperature (hence, in both boxes) the chemical transitions between $C$ and $A$  are faster than the ones between $B$ and $A$. For simplicity,  we henceforth choose $\Delta \epsilon_C = 0$.

We are interested in the probability of finding the system in the lowest energy states, $B$ and $C$, at stationarity. 
In the following, $P(B)$ is identified as $P(B_1) + P(B_2)$, and analogously for $P(C)$. When equilibrium conditions are met, (namely $d_A = d_B = d_C = 0$ and/or $\Delta T = 0$), the system asymptotically converges to $P^{eq}_i(C) = P^{eq}_i(B) > P^{eq}_i(A)$ in each box and consequently $P^{eq}(B) = P^{eq}(C) > P^{eq}(A)$ overall. 
In non-equilibrium conditions the picture dramatically changes, because the energy symmetry between states $B$ and $C$ is kinetically broken. 
In order to emphasize the role of the barrier $\Delta \epsilon$, we set all the transport rates to be equal, $d_A = d_B = d_C = d$. In this simple setting, away from equilibrium the state with the lowest energy barrier, $C$ in this case, is the most populated at steady-state in the presence of a temperature gradient. This is quantified by the ratio between the probabilities of the $C$ and $B$ states,  $R_{CB} = P(C)/P(B)$, whose logarithm can be interpreted as the effective  stabilization energy of $C$  relative to $B$ (Fig.\ref{fig:1}B). 
$R_{CB}$ is always greater than $1$, and it is a monotonously increasing function of $d$. It reaches its maximum value in the $d \to \infty$ limit, \textit{i.e.} when diffusion between the two boxes is much faster than all other processes in the system.
In this limit it is possible to find the analytic expression of $R_{CB}$ for an arbitrary number $n$ of boxes:
\begin{equation}
\lim_{d\to \infty} R_{CB} = \frac{\hat{k}_{B \to A}   \hat{k}_{A \to C}}{\hat{k}_{C \to A} \hat{k}_{A \to B}} 
\label{RCB}
\end{equation}
with $\hat{k}_{X \to Y} = \sum_i^n k_{X_i \to Y_i}$. 

The simple model that we have proposed here provides a clear example of kinetic symmetry-breaking due to the energy barriers, which is effective only in a non-equilibrium scenario \cite{astumkin}. In particular, the state which is more favorable away from  equilibrium, $C$, participates in the reactions that, according to (\ref{barriers}), are the fastest.
The role of $\Delta \epsilon$ in the selection process is revealed in a small $\Delta T$ expansion of  Eq. \eqref{RCB}:
\begin{equation}
R_{CB} = 1 + \frac{\Delta E \Delta T^2}{4 T_2^4} \Delta \epsilon + \mathcal{O}(\Delta T^3)
\label{RCBDT}
\end{equation}
As expected, the zeroth order is equal to $1$, since at equilibrium the states $B$ and $C$ are equally populated. Furthermore, the first order term vanishes because the selection of the fastest state cannot depend on the direction of the temperature gradient. 

\subsection{State selection is governed by dissipation.}

\begin{figure}[th]
\centering
\includegraphics[width=1\columnwidth]{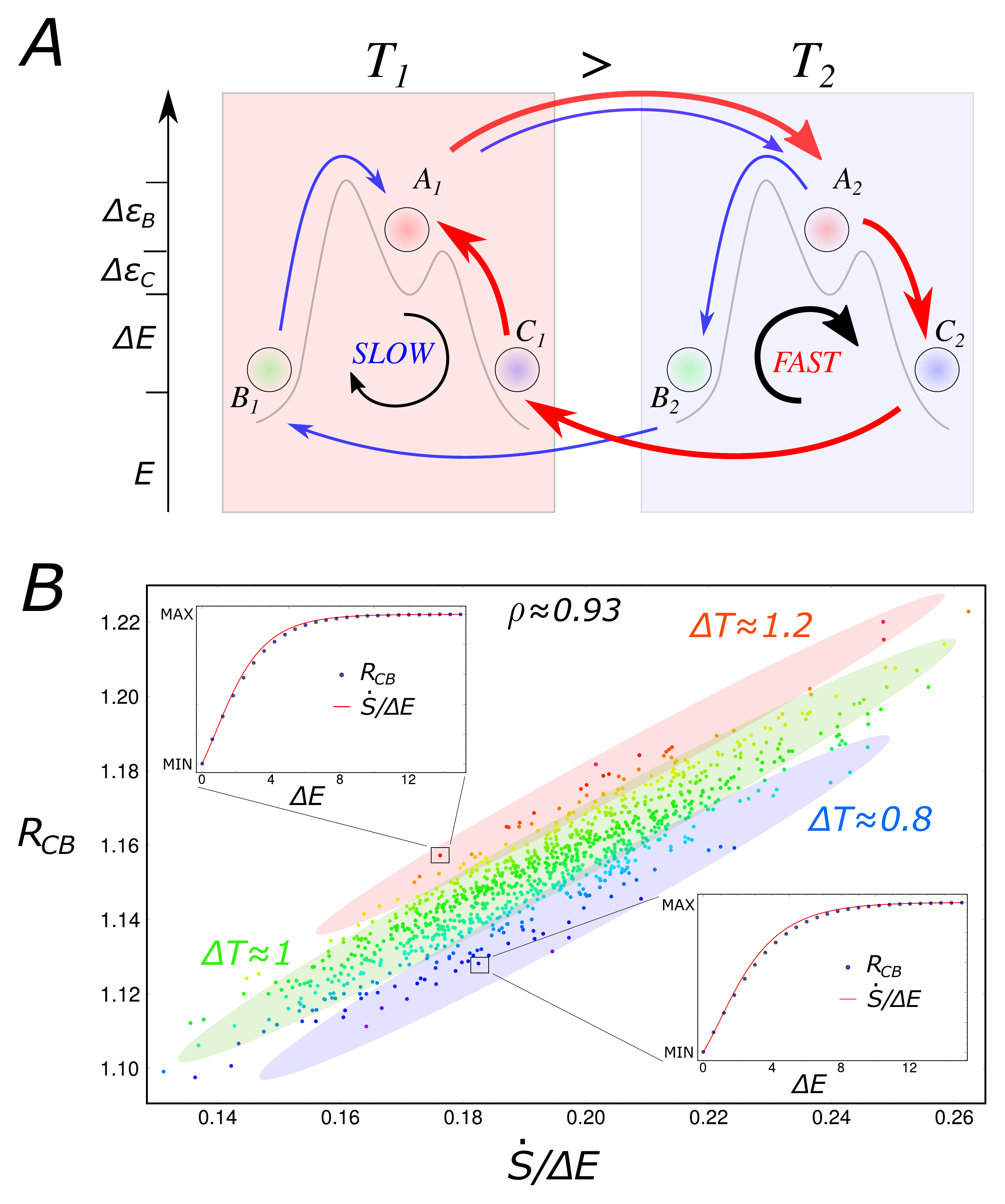}
\caption{A) Diffusive cycles convert thermal energy into chemical energy. The direction and thickness of each arrow represent respectively direction and intensity of the net probability flux between two states. B) Correlation between $R_{CB} = P(C)/P(B)$ and $\dot{S}/\Delta E$, which is the steady state entropy production divided by the characteristic energy scale of the system, for different values of $\Delta E$ and $\Delta T$. Here, $\Delta \epsilon = 1$, $T_2 = 1$, $d\to +\infty$, both $\Delta E$ and $\Delta T$ have been drawn from a normal distribution of mean $1$ and variance $0.1$. Identical values of the gradient correspond to the same color. $\rho$ is the correlation coefficient. We report the approximate average $\Delta T$ among the values contained in each shaded area. \textit{Insets} - Setting $\Delta T = 1.2$ (top) and $\Delta T = 0.8$ (bottom), we show that $R_{CB}$ and $\dot{S}/\Delta E$ exhibit the same behavior as a function of $\Delta E$, when plotted within the same range. The Boltzmann coefficient has been taken equal to $1$ for simplicity.}
\label{fig:2}
\end{figure}

An intuitive grasp of the mechanism leading to selection of the fastest state can be provided by Fig.\ref{fig:2}A, where the direction of the currents have been highlighted. Thermal energy is converted into chemical energy, namely excess of $C$ over $B$, through diffusive cycles taking place in the system. 
Particles are heated up in the hot box ($B$ and $C$ toward $A$), thus absorbing heat, whereas they rellax ($A$ to $B$ and $C$) in the cold box, thus releasing heat. This unbalance generates a current of $A$ from the warm to the cold box, where it splits preferentially along the faster decay path, that is, toward $C$, before being transported back into the hot box. This cycle is thus driven by the constant absorption and dissipation of energy, which is related to \textit{entropy production} \citep{schnakenberg, bamos2}:
\begin{eqnarray}
\dot{S} &=& \sum_{i=1}^2 \sum_{X=B,C} J_{A_i\to X_i} \ln\frac{k_{A_i\to X_i}}{k_{X_i\to A_i}}= \nonumber \\
&=& \Delta E \frac{\Delta T}{T_1 T_2} \left( J_{A_2 \to B_2} + J_{A_2 \to C_2} \right)
\label{Sss}
\end{eqnarray}
where $J_{A_i\to X_i} = k_{A_i\to X_i} P(A_i) - k_{X_i\to A_i} P(X_i)$ is the flux from $A_i$ to $X_i$, with $i$ indicating the box. We used $J_{A_1 \to X_1} = -J_{A_2 \to X_2}$ (Fig. \ref{fig:2}), and the contributions from the inter-box currents vanish because the rates in the two directions are equal. $\dot{S}$ is positive because the currents flow away from A at the colder temperature ($T_2$) and toward A at the warmer one ($T_1$).

Expanding Eq. \eqref{Sss} up to the second order in $\Delta T$, and using Eq. \eqref{RCBDT}, we have:
\begin{equation}
R_{CB} \simeq 1 + \frac{\dot{S}}{\Delta E} \frac{1}{P^{eq}(B)} \frac{\Delta \epsilon}{1+e^{\Delta \epsilon/T_2}}
\end{equation}
Despite the validity of this formula only for small gradients and fast diffusion, it suggests a correlation between $R_{CB}$, which quantifies \textit{selection}, and $\dot{S}/\Delta E$, which is related to dissipation in the system.

In Fig.\ref{fig:2}B, we show that indeed $R_{CB}$ and $\dot{S}/\Delta E$ are highly correlated for a set of (random) thermal gradients $k_B \Delta T = k_B (T_1 - T_2)$ and values of the typical energy scale $\Delta E$.
Here it is clear that the gradient $\Delta T$ quantifies the available (thermal) energy driving the selection of the fastest state $C$ through dissipation. Indeed, as $\Delta T$ increases, the probability of escaping from $B$, diffusing, and populating $C$ increases as well.
In the SI we show how the correlation changes for different values of the energy barrier $\Delta \epsilon$. Remarkably, fixing the thermal gradient, $R_{CB}$ is always strongly correlated with the steady-state entropy production as a function of the energy $\Delta E$ (Insets of Fig.\ref{fig:2}B). 

\subsection{Characteristic lengthscale for selection.}

Extending this two-box model to a thermal gradient in continuous space is of course more realistic, and reveals further features that are inaccessible to the discrete box description.
In continuous space (say, $x\in [0,1]$), the system evolves according to the differential Chapman-Kolmogorov equation \cite{gardiner}:
\begin{eqnarray}
\partial_t p_X(x) &=& \sum_{Y} \left( k_{Y \to X}(x) p_Y(x) - k_{Y \to X}(x) p_X(x) \right) + \nonumber \\
&\;& + ~D_X \partial_x^2 p_X(x)
\label{Ch-Kol}
\end{eqnarray}
where $X,Y = A,B,C$. We impose no-flux boundary conditions, \textit{\textit{i.e.}} $\partial_x p_X(0) = \partial_x p_X(1) = 0$. 
In (\ref{Ch-Kol}), the Laplacian captures diffusion while the part involving discrete transitions captures the chemical reactions between species, which are governed by rates analogous to the ones introduced before:
\begin{eqnarray}
k_{A \to B}(x) = e^{(E_A - E_B)/k_B T(x)} k_{B \to A}(x) \nonumber \\
k_{A \to C}(x) = e^{(E_A - E_C)/k_B T(x)} k_{C \to A}(x)
\label{ratesDiff}
\end{eqnarray}
with the additional condition on the energy barriers:
\begin{eqnarray}
k_{C \to A}(x) = e^{\Delta \epsilon /k_B T(x)} k_{B \to A}(x)\quad .
\end{eqnarray}
Also in this case, the transport coefficient is the same for all states:  $D_X \equiv D$, $\forall X$.
In what follows $P(X) = \int dx ~p_X(x)$ (note that we use $p$ for the space dependent distribution, and $P$ to indicate their integral over space). 

Although it is difficult to solve Eqs.~\eqref{Ch-Kol} analytically for any value of the parameters, approximate solutions can be worked out in selected cases.
The limit of large diffusion ($D \to \infty$), which is analogous to the case of infinitely fast transport between the two boxes analyzed above, can be tackled using the standard approach of time-scale separation \cite{celani, gardiner}. 
To the leading orders in $1/D$, the solution is uniform in space, and $R_{CB}$ is the same as in (\ref{RCB}), with 
$\hat{k}_{X \to Y} = \int dx ~k_{X \to Y}(x)$.

The case of a linear temperature gradient $T(x) = T_0 + \Delta T \cdot x$ can also be analytically explored for small $\Delta T$. Expanding all rates and probabilities in powers of $\Delta T$ as
\begin{eqnarray}
&& k_{X \to Y} = \sum_n \frac{1}{n!} x^n \Delta T^n \partial_T^n k_{X\to Y}|_{\Delta T = 0} \nonumber \\
&& p_X(x) =  \sum_n \Delta T^n p_X^{(n)}(x)\quad ,
\end{eqnarray}
inserting them in (\ref{Ch-Kol}) and solving it order by order it is easy to obtain at 0th order
\begin{equation}
p^{(0)}_B = p^{(0)}_C = \frac{e^{\Delta E/k_B T_0}}{2 e^{\Delta E/k_B T_0} + 1}
\end{equation}
which is the equilibrium solution for $\Delta T = 0$.

Up to second order, $R_{CB}$ is
\begin{equation}
R_{CB} = 1 +  \frac{\Delta T^2}{2 P_B^{(0)}} \left(P_{C}^{(2)}-P_{B}^{(2)}\right) 
\end{equation}
where $P^{(n)}_X$ is defined as the integral of $p^{(n)}_X(x)$ over the whole domain. 
After a further expansion in $\Delta \epsilon$, \textit{i.e.} the symmetry between $B$ and $C$ is only infinitesimally broken by the kinetics, we obtain
\begin{equation}
P^{(2)}_C - P^{(2)}_B = \frac{\Delta E}{T_0^4} P^{eq} L_s^2 \left( 1 - 2 L_s \tanh \left( \frac{1}{2 L_s} \right) \right) \Delta \epsilon
\label{pcmenopb}
\end{equation}
with $L_s = \sqrt{D/(k_{B \to A} + 2 k_{A \to B})}$ and $P^{eq} = P^{(0)}_B = P^{(0)}_C$. This difference is always positive, implying that states participating in fast reactions are always favored, and it vanishes when $D \to 0$, as expected because the system locally relaxes at equilibrium.
In particular, $L_s$ represents a typical length-scale that can be interpreted as the space traveled by the system between two state transitions, namely the distance below which the system can absorb and dissipate energy, thus setting a lengthscale for dissipation-driven selection.

When $\Delta T \to \infty$, all states tends to be equally populated, \textit{i.e.} $R_{CB} \to 1$, abolishing chemical selection. Since $R_{CB} = 1$ also when $\Delta T =0$, and is always positive, it must have a maximum at a given $\Delta T^*$, as we numerically show in the SI, suggesting that maximal selection would stem from a fine tuning of the parameters of the chemical network for any given $\Delta T$.

\subsection{Non-trivial selection for more complex reaction-network topologies.}

How do the features of a simple three-states system extend to more complex network topologies? Here we study a chain of connected chemical reactions in an energy landscape, looking at the propagation of the local selection process along the chain, eventually leading to runaway and/or localization phenomena in the population of states. 

\begin{figure}[t]
\centering
\includegraphics[width=\columnwidth]{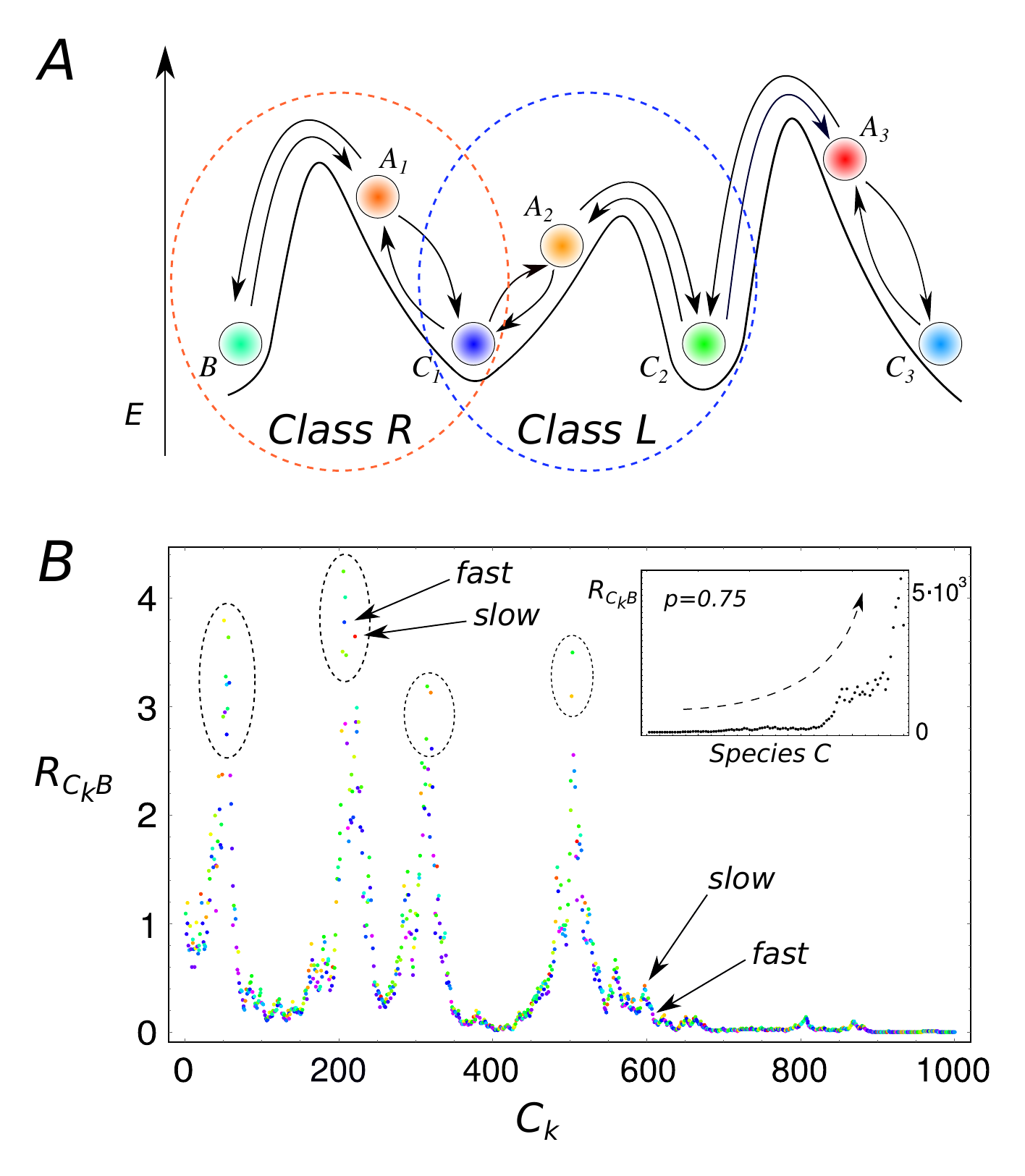}
\caption{A) Chain of three concatenated three-states chemical networks, each similar to the one in Fig.~\ref{fig:1}A. The orange circle indicates a subsystem belonging to the class $R$, with the fast transition on the right branch, while the blue circle indicates a subsystem whose fast transition is on the left branch (class $L$).  B) $R_{C_kB} = P(C_k)/P(B)$ as a function of the states $C_k$. The selection of states does not depend only on their transitions being fast or slow with respect to the neighboring reactions, but on all rates of the network. Here $k_B = 1, T = \Delta T = 1$, $\Delta E_k \sim U([1,10])$ and $\Delta \epsilon_k \sim U([0.1,2])$, where $U$ is the uniform distribution. Each subsystem belongs to class $R$ with probability $p=0.5$, and to class L with probability $1-p$. \textit{Inset} - $R_{C_kB}$ as a function of the species $C_k$ for the same parameters as in the main panel, but $p=0.75$. The predominance of subsystems belonging to the class $R$ leads to a ``directional" exponential growth.}
\label{fig:3}
\end{figure}

We focus to the two-box scenario, which, as shown above, recapitulates most of the dissipation-driven selection phenomenon while being easier to analyze, in the limit of infinitely fast transport between the boxes. We consider a reaction network as the one sketched in Fig.~\ref{fig:3}A, which can also diffuse between two boxes at different temperatures as in Fig.\ref{fig:1}. We can distinguish two different classes of three-state subsystems, with the faster reaction either on the right branch (henceforth indicated as $R$, encircled by an orange dashed line in Fig.~\ref{fig:3}A), analogously to the three state system depicted in Fig.~\ref{fig:1}A, or on the left ($L$, encircled by a blue dashed line  in Fig.~\ref{fig:3}A). All lower-energy states have the same energy, while the high energy state in each subsystem is characterized by a different energy $\Delta E_i$, and a different barrier $\Delta \epsilon_i$, mimicking the presence of a non-trivial underlying energy landscape.

We have already computed $R_{CB}$ in Eq.~\eqref{RCB}, in the limit of infinitely fast diffusion. It quantifies the ratio between the population of two adjacent states, the fast over the slow one. It is possible to see from the Master Equation for the whole system in Fig.~\ref{fig:3}A, that the same relation holds between any two adjacent states in each subsystem. Since we want to compute the population of each single species along the chain, we use $B$ as our reference state. The ratio between $P_{C_k}$ and $P_B$ is: 
\begin{equation}
R_{C_k B} = R_{C_1 B} \prod_{l=2}^k R_{C_lC_{l-1}}
\label{RCBchain}
\end{equation}
If there are $n_k^{(L)}$ subsystems belonging to the class $L$, and $n_k^{(R)} = k - n_k^{(L)}$ subsystems to the class $R$, then Eq. \eqref{RCBchain} becomes:
\begin{equation}
R_{C_kB} = \prod_{i=1}^{n^{(R)}} R_i^{(R)}(\Delta E_i, \Delta \epsilon_i) \prod_{i=1}^{n^{(L)}} R_i^{(L)}(\Delta E_i, \Delta \epsilon_i)
\end{equation}
with
\begin{gather}
R_i^{(R)} = 1 + \tanh\left( \frac{\Delta E_i \Delta T}{2 T (T + \Delta T)} \right)\tanh\left( \frac{\Delta \epsilon_i \Delta T}{2 T (T + \Delta T)} \right) \nonumber \\
R_i^{(L)} = \left( R_i^{(R)}(\Delta E_i, \Delta \epsilon_i) \right)^{-1}
\label{RCBLR}
\end{gather}

To simulate a generic chain of chemical reactions, we assign each subsystem to class $L$ with probability $p$, and to class $R$ with probability $q=1-p$. We then draw $\Delta E_i$ and $\Delta \epsilon_i$ from two distributions, $P(\Delta E)$ and $P(\Delta \epsilon)$ respectively (details in the caption of Fig.\ref{fig:3}). As we can see from Fig.~\ref{fig:3}B, even in the simple case in which $p=q=1/2$, and both distributions are uniform, a localization phenomenon in the population of the states can spontaneously arise, where the \textit{favorability} of an individual state does not depend only on its fast/slow status with respect to the adjacent reactions, but depends instead on the full path of reactions connecting it to the reference state, and hence on the full energy landscape.
If all the fast reactions are on the same side of each three-state subsystem (all reactions of type $R$ or of type $L$), the population of states $C_k$ can become exponentially different from the one of $B$, as highlighted in the inset of Fig.\ref{fig:3}B. 

Also in this case, the selection for the most probable states is determined by dissipation. The argument outlined for the simple three-state system can be easily generalized in the case of infinitely fast transport between the boxes: $R_{C_k B}$ simply corresponds to the product of all the transition rates directed from $B$ to $C_k$ belonging to the path connecting the two, divided by the same product in the opposite direction. As a consequence, the states that will eventually be the most populated ones (with respect to a reference state $B$) are those whose connecting path to $B$ have the fastest dissipation. 
However, when the topology is further complicated, several distinct paths can connect the same pair of states, and \textit{all} the transition rates will eventually contribute to determine a ranking for steady-state populations. In this case the determination of the fastest dissipating states becomes difficult, and we leave for future works the development of an efficient technique to tackle this problem.

\subsection{Emergence of thermophoresis-like behavior.}

So far we have assumed that all the species move between the boxes (or diffuse in space) at the same rate, and as a consequence the probability to be in each box, summed over the different states, is always equal to $1/2$ (or uniform in continuous space). Although relaxing this hypothesis does not significantly change the overall picture of dissipation-driven selection, a novel phenomenon appears, that we are compelled to report for its potential implications: we find that, even in the simple two-box scenario, there is an accumulation of the population in one of the boxes.  The description of this effect is surprisingly similar to thermophoresis, which refers to the {\color{black}accumulation of molecules on either the cold or warm side in presence of a thermal gradient. Mathematically, at stationarity, thermophoresis is usually described through a diffusive equation \cite{duhr, piazza}:
\begin{equation}
\nabla c = - S_T c \nabla T
\label{ST}
\end{equation}
where $c$ is the concentration of particles, and $S_T$ is the so-called Soret coefficient, which can be positive or negative. Even if extensively described through effective equations, a microscopic understanding of this behaviour is still lacking \cite{platten, soret2, soret1}. The present approach might serve as a complementary perspective for this intriguing phenomenon.

To fix the ideas, consider the discrete-state system sketched in Fig.\ref{fig:1}A. We consider the ratios $d_B/d_A$ and $d_C/d_A$ as measures of the unbalance of transport properties of different species. The probability of being in box $i$ is $P_i = P(A_i)+P(B_i)+P(C_i)$. In a discrete box scenario, Eq. \eqref{ST} can be rewritten as:
\begin{equation}
\Delta P = P_2 - P_1 = - S_T \left( \frac{P_1 + P_2}{2} \right) \Delta T
\end{equation}
Since for infintely fast transport the system will end up equally populating both boxes, we need to consider finite transport. 
$\Delta P = P_2 - P_1$ is represented in Fig.\ref{fig:5} as a function of $d_B/d_A$, for two different choices of $ d_C/d_A$, and for different values of $\Delta T$. Clearly, in the absence of a thermal gradient there is no thermophoresis, while a difference between $T_1$ and $T_2$ induces an accumulation of particles on the warm or cold side.
When the transport coefficients are small compared to all the other transition rates in the system, the Soret coefficient can be estimated to be equal to:
\begin{equation}
S_T = \frac{(2 - \frac{d_B + d_C}{d_A}) ~\Delta E ~e^{\Delta E/k_B T_2}}{(1 + 2 e^{\Delta E/k_B T_2}) (\frac{d_B + d_C}{d_A} e^{\Delta E/k_B T_2}  + 1) k_B T_2^2}
\label{Soretcoeff}
\end{equation}
As can be seen from (\ref{Soretcoeff}), the sign of $S_T$ depends on the values of the transport coefficients of the different states, and it thus inextricably links transport to the internal kinetics in ``chemical" space. Indeed, even a simple two-state system exhibits thermophoresis, as long as the two states have different transport coefficients (see SI).

In line with the leit-motif of this work, we highlight here that thermophoresis can again be seen as a selection process in position, rather than in state, space.  It is driven by the dissipation of thermal energy, and the kinetic symmetry-breaking is induced by the asymmetry of transport rates.

\begin{figure}[t]
\centering
\includegraphics[width=1 \columnwidth]{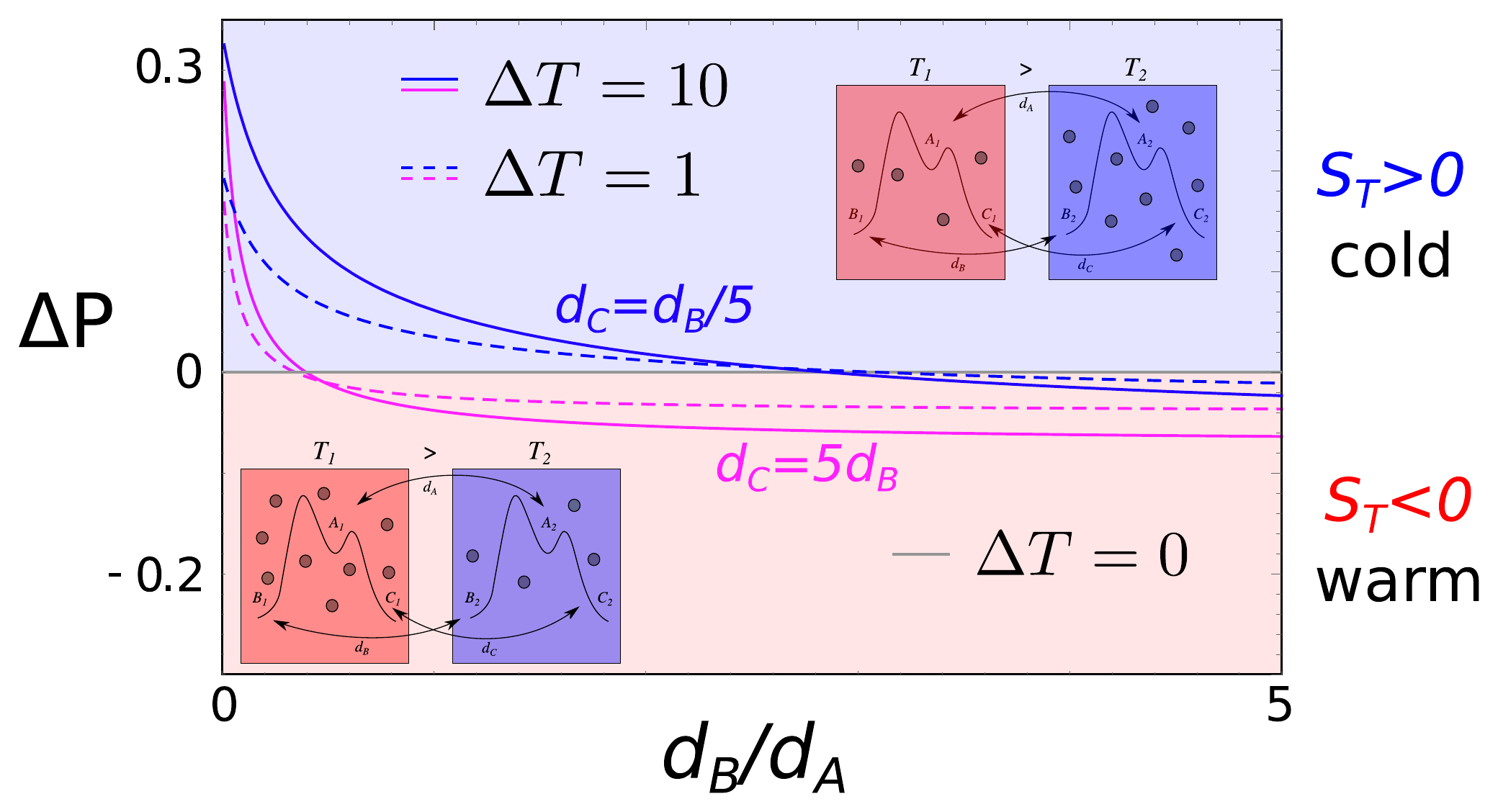}
\caption{Difference between the probability of being in each box, $\Delta P = P_2 - P_1$, as a function of $d_B / d_A$, for $d_C = d_B / 5$ (blue curves) and $d_C = 5 d_B$ (magenta curves). Different values of the gradient $\Delta T$ are shown.  When $\Delta P$ is positive, the particles (independently of the species) accumulate on the cold side (blue box), and the Soret coefficient $S_T$ is positive. On the contrary, the particles are more abundant in the warm side (red box) for negative $\Delta P$, corresponding to negative values of $S_T$. In this example, we set $T=1, \Delta E = 0.1, \Delta \epsilon = 2$ and $d_A = 1$. $k_B$ has been taken equal to $1$ for simplicity.}
\label{fig:5}
\end{figure}

\section{Discussions}

Non-equilibrium conditions can trigger stabilization effects in molecular systems \citep{sassi, assenza}. In a similar fashion, here we have shown that high-energy states can be stabilized out-of-equilibrium, by continuously dissipating energy supplied from an external source, a temperature gradient in our case. States participating to the fastest reaction pathways will be the most populated ones at steady-state. Hence, the core ingredient is the breakdown of kinetic symmetry in the reaction rates: while at equilibrium the  energies are the only relevant quantities, away from equilibrium the kinetics plays a fundamental role. Here we have proposed simple reaction networks that could be investigated to reveal how selection and dissipation are intimately related. Furthermore, because of their simplicity, these models can be analytically and numerically solved and, importantly, are amenable of experimental validation.
As a byproduct of our study, we have presented a thermophoresis-like behaviour emerging as a spatial selection process. This is induced, again, by kinetic symmetry-breaking, in this case in the diffusion coefficients of different states.
It is also worth noting that the relation between selection and dissipation stems from the thermodynamic necessity to transport heat from the warm to the cold side of the system. In this respect, selection becomes a necessary consequence of thermodynamics.

From a broader perspective, this work could provide a novel framework to develop schemes aimed at explaining the sustained abundance of otherwise only metastable molecules, which are necessary intermediates for the spontaneous synthesis of more complex macromolecules that, in turn, could lead to the first replicators. In this respect, we are convinced that our result is a first step in connecting the \textit{origin of life} problem into the physical questions of what is possible in non-equilibrium conditions, and what are the basic microscopic (molecular) rules governing the emergent phenomena. 
\\

\section*{Acknowledgments}

We acknowledge A. Maritan, F. Stellacci, C. Jarzynski and D. Astumian for useful discussions, M. A. Younan for the help in developing the expansion for small value of the gradient in the continuous-space description and for insightful observations, V. Ouazan for thoughful comments, and F. Piazza for the careful reading of the manuscript and inspiring remarks.


\newpage
\begin{widetext}

\appendix

\section*{\large{Supplementary Information for ``Dissipation-driven selection in non-equilibrium chemical networks"}}

\maketitle

\section*{\MakeUppercase{Three chemically interacting species with diffusion}}

Here, we present some mathematical details about the continuous version of the discrete-state model. The dynamics, presented in the main text, follows a differential Chapman-Kolmogorov equation:
\begin{equation}
\partial_t p_X(x) = \sum_{Y} \left( k_{Y \to X}(x) p_Y(x) - k_{Y \to X}(x) p_X(x) \right) + D_X \partial_x^2 p_X(x)
\label{cha-kol}
\end{equation}
where $X,Y = A,B,C$ and $x\in [0,1]$ with boundary conditions $\partial_x p_X(0) = \partial_x p_X(1) = 0$, i.e. no flux at the boundary. For sake of simplicity, we set $D_A = D_B = D_C = D$, as for the discrete-state outlined in the main text. Let us recall that the rates governing the transitions are:
\begin{eqnarray}
k_{A \to B}(x) = e^{(E_A - E_B)/k_B T(x)} k_{B \to A}(x) \nonumber \\
k_{A \to C}(x) = e^{(E_A - E_C)/k_B T(x)} k_{C \to A}(x) 
\end{eqnarray}
with the additional condition on the energy barriers:
\begin{eqnarray}
k_{C \to A}(x) = e^{\Delta \epsilon /k_B T(x)} k_{B \to A}(x)
\end{eqnarray}
where the barrier $\Delta \epsilon > 0$, so that the transition involving $A$ and $C$ is faster than the one involving $A$ and $B$.

Here, the probability to find the system in a given state $X$, independently of the position, is:
\begin{equation}
P(X) = \int dx p_X(x) \qquad X = A,B,C
\end{equation}
We will use small $p_X(x)$ for space-dependent probabilites, and capital $P(X)$ for their space integrated counterpart.

We have already seen in the main text that, in equilibrium conditions, the system asymptotically converges to $P^{eq}_B = P^{eq}_C > P^{eq}_A$. If there is no temperature gradient, the system relaxes to the Boltzmann distribution determined by its global temperature $T$ for all the species. On the contrary, in case of no diffusion, the probability for each species is Boltzmann distributed as a function of the local temperature $T(x)$.

\subsection*{Limit $D \to \infty$}

Following the standard approach of the time-scale separation \cite{gardiner, celani}, it is possible to define a set of effective transition rates, $\hat{k}_{X\to Y}$, such that the stationary solution, in the limit $D \to \infty$, is given by the one of a discrete-state system subject to these $\hat{k}_{X\to Y}$, without diffusion. Naively speaking, dividing the space in $n$ boxes, we are imaging that for $D \gg k_{X \to Y}$ each species will feel the contribution from each box at the same time, performing then transitions at an effective rate given by:
\begin{equation}
\hat{k}_{X \to Y} = \int dx \mu(x) k_{X \to Y}(x)
\end{equation}
where $\mu(x)$ is the probability distribution satisfying $D_X \partial_x^2 P(X(x)) = 0$, which is uniform in $x$. Then, the solution for each species can be readily found:
\begin{eqnarray}
P(A) = \int dx ~p_A(x) &=& \frac{\hat{k}_{B \to A} \hat{k}_{C \to A}}{\hat{k}_{B \to A} \hat{k}_{A \to C} + \hat{k}_{C \to A} \hat{k}_{A \to B} + \hat{k}_{C \to A} \hat{k}_{B \to A}} \nonumber \\
P(B) = \int dx ~p_B(x) &=& \frac{\hat{k}_{C \to A} \hat{k}_{A \to B}}{\hat{k}_{B \to A} \hat{k}_{A \to C} + \hat{k}_{C \to A} \hat{k}_{A \to B} + \hat{k}_{C \to A} \hat{k}_{B \to A}} \nonumber \\
P(C) = \int dx ~p_C(x) &=& \frac{\hat{k}_{B \to A} \hat{k}_{A \to C}}{\hat{k}_{B \to A} \hat{k}_{A \to C} + \hat{k}_{C \to A} \hat{k}_{A \to B} + \hat{k}_{C \to A} \hat{k}_{B \to A}}
\end{eqnarray}

This solution gives exactly the same result as in Eq.~$(5)$ of the main text.

\subsection*{Perturbation theory approach}

It is interesting to explore in details the limit of small gradients for the continuous case. The perturbative approach that we are going to mention is valid for any form the transition rates. Then, in order to keep the analysis as general as possible, we simply say that:
\begin{equation}
\kappa(x) = \frac{k_{A \to B}(x)}{k_{B \to A}(x)} = \frac{k_{A \to C}(x)}{k_{C \to A}(x)} > 1 \qquad \qquad \alpha(x) = \frac{k_{A \to C}(x)}{k_{A \to B}(x)} > 1
\end{equation}
Note that here $\alpha$ plays the same role as the energy barrier $\Delta \epsilon$. Eventually, we restrict ourselves to the physical choice of the Arrenhius' form when comparing these results with the one for the discrete case.

The temperature dependence appears only in the transition rates. Then, assuming that the temperature gradient is constant, \textit{i.e.} $T(x) = T_0 + \Delta T \cdot x$, we can expand each of them as follows:
\begin{equation}
k_{X \to Y} = \sum_n \frac{1}{n!} x^n \nabla T^n \partial_T^n k_{X\to Y}|_{\nabla T = 0}
\end{equation}
From now on we will not write explicitly the fact that the expansion coefficient are evaluated at $T(x) = T_0$.
 
We can perform the same expansion also on the probabilities, as $p_X(x) = \sum_n \nabla T^n p^{(n)}_X(x)$. Substituting this into the dynamical equation, and using the fact that $p_A(x) + p_B(x) + p_C(x)$ is uniform at stationarity (since $D_A = D_B = D_C = D$), and equal to $1$ for simplicity in a $1D$ box of unitary length, we get that the following $n$-th order set of equations for the steady state has to be fulfilled:
\begin{equation}
D \partial_x^2 p_X(x) = \sum_{m+l = n} \frac{1}{m!} x^m \bigg( \big( \partial_T^m k_{A \to X} + \partial_T^m k_{X \to A} \big) p^{(l)}_B(x) + \partial_T^m k_{A \to X} p^{(l)}_C(x) \bigg) - \frac{1}{n!} x^n \partial_T^n k_{A \to X}
\label{norder}
\end{equation}
for $X = B,C$ only, with the boundary conditions $\partial_x p^{(n)}_X(x) = 0$ at $x = 0,1$. For sake of simplicity, from now on we use the following positions: 
\begin{eqnarray}
\mathcal{K}^{(X)}_n \equiv \partial_T^n k_{A\to X} &\qquad & \qquad k^{(X)}_n = \partial_T^n k_{X\to A}
\end{eqnarray}
Then we can solve Eq.~\eqref{norder} in terms of the probability coefficients $p^{(n)}_X(x)$. It is worth showing the zeroth and first order corrections separately, since they elucidates some features of the system.

\subsubsection*{Zeroth order solution}

At the zeroth order, as reported also in the main text, we obtain:
\begin{equation}
p^{(0)}_B(x) = p^{(0)}_C(x) = \frac{\kappa(x)}{2 \kappa(x) + 1}
\end{equation}
Then $P^{(0)}(B) - P^{(0)}(C) = 0$ at the lowest order, meaning that, as expected, a thermal gradient is needed to allow reaching a non-equilibrium stationary state. Here $P^{(n)}(X)$ is the integral of $p^{(n)}_X(x)$ over the whole domain.

\subsubsection*{First order solution}

In order to get the first order solution, we have to solve Eq.~\eqref{norder} with $n=1$. For sake of clarity, let us introduce the following constants:
\begin{equation}
U_n^{(X)} = \frac{K^{(X)}_n+k^{(X)}_n}{D} \qquad \qquad V_n^{(X)} = \frac{K^{(X)}_n}{D}
\end{equation}
Rewriting the dynamical equation in terms of these quantities and $P_0(X(x))$, by direct integration, we get:
\begin{equation}
p^{(1)}_X(x) = 2 \left( M^+_X - M^-_X \right) x + 2 M^+_X L^+ \frac{\sinh \left( \frac{1-2x}{2L^+} \right)}{\cosh \left( \frac{1}{2L^+} \right)} - 2 M^-_X L^- \frac{\sinh \left( \frac{1-2x}{2L^-} \right)}{\cosh \left( \frac{1}{2L^-} \right)}
\end{equation}
where, with $X$ always intented as $B,C$:
\begin{eqnarray}
W^{(X)} &=& U_1^{(X)} P_0(B(x)) + V_1^{(X)} P_0(C(x)) - V_1^{(X)} \nonumber \\
J &=& \sqrt{\left( U_0^{(C)} - U_0^{(B)} \right)^2 + 4 V_0^{(B)} V_0^{(C)}} \nonumber \\
L^{\pm} &=& \sqrt{\frac{2}{U_0^{(C)} + U_0^{(B)} \pm J}} \nonumber \\
M^{\pm}_B &=& \mp \frac{(L^\pm)^2}{2} \left( W^{(B)} \frac{\mp U_0^{(C)} \pm U_0^{(B)}+J}{2J} \pm W^{(C)} \frac{V_0^{(B)}}{J} \right) \nonumber \\
M^{\pm}_C &=& \mp \frac{(L^\pm)^2}{2} \left( W^{(C)} \frac{\mp U_0^{(C)} \pm U_0^{(B)}+J}{2J} \pm W^{(B)} \frac{V_0^{(C)}}{J} \right)
\label{vars}
\end{eqnarray}
Considering the physical situation in which we have the Arrhenius's form for the transition rates \cite{raz, jarz, bamos1}:
\begin{equation}
p^{(1)}_C(x) - p^{(1)}_B(x) = \left( M^+_B - M^+_C \right) \left( 2 L^- \frac{\sinh \left( \frac{1-2x}{2L^-} \right)}{\cosh \left( \frac{1}{2L^-} \right)} - 2 L^+ \frac{\sinh \left( \frac{1-2x}{2L^+} \right)}{\cosh \left( \frac{1}{2L^+} \right)} \right)
\end{equation}
Most importantly, this expression is symmetric around $x = 1/2$ in our domain of length $1$. This means that the unbalance between state $C$ and $B$ is independent of the sign of the gradient.

\subsubsection*{Second order solution for $\Delta \epsilon \to 0$}

The $n$-th order solution can be readily found by direct integration of the $n$-th order dynamical equations, getting:
\begin{equation}
P^{(n)}(C) - P^{(n)}(B) = \frac{1}{U_0^{(B)} U_0^{(C)} - V^{(B)}_0 V^{(C)}_0} \bigg( (U_0^{(B)} + V_0^{(B)}) \int_0^1 dx ~g_n(x) - (U_0^{(C)} + V_0^{(C)}) \int_0^1 dx ~f_n(x) \bigg)
\label{nth}
\end{equation}
It is clear that this expression is well defined once one knows the solution at the lower orders.

The second order solution can be obtained by specializing Eq.~\eqref{nth} with $n=2$ and specifying the functional form of $f_2(x)$ and $g_2(x)$ as:
\begin{eqnarray}
f_2(x) = \frac{x^2}{2} \left( U_2^{(B)} p^{(0)}_B(x) + V_2^{(B)} p^{(0)}_C(x) - V_2^{(B)} \right) + x \left( U_1^{(B)} p^{(1)}_B(x) + V_1^{(B)} p^{(1)}_C(x) \right) \nonumber \\
g_2(x) = \frac{x^2}{2} \left( U_2^{(C)} p^{(0)}_B(x) + V_2^{(C)} p^{(0)}_C(x) - V_2^{(C)} \right) + x \left( U_1^{(C)} p^{(1)}_B(x) + V_1^{(C)} p^{(1)}_C(x) \right)
\end{eqnarray}

We note that all the terms that do not scale with the diffusion have to cancel out. Indeed, when $D \to 0$, there is no difference between $P(C)$ and $P(B)$, as they have the same energy and the system relax to the Boltzmann distribution. Then, after some calculations, taking, for sake of simplicity, $\Delta E \equiv E_A - E_B$ and $E_B = E_C$, we obtain:
\begin{eqnarray}
P^{(2)}(C) - P^{(2)}(B) &=& \frac{\Delta E}{2 T_0^4} \frac{V_0^{(B)}}{U_0^{(B)} + V_0^{(B)}} \left( \frac{U_0^{(C)} - U_0^{(B)} - 2 V_0^{(C)}}{J} + 1 \right) \Delta \epsilon ~(L^-)^2 \left( 1 - 2L^- \tanh \left( \frac{1}{2 L^-} \right) \right) + \nonumber \\
&\;& - ~\frac{\Delta E}{2 T_0^4} \frac{V_0^{(B)}}{U_0^{(B)} + V_0^{(B)}} \left( \frac{U_0^{(C)} - U_0^{(B)} - 2 V_0^{(C)}}{J} - 1 \right) \Delta \epsilon ~(L^+)^2 \left( 1 - 2L^+ \tanh \left( \frac{1}{2 L^+} \right) \right)
\label{2order}
\end{eqnarray}

In order to interpret this formula, we consider the limit $\Delta \epsilon \to 0$, i.e. the energy barrier discriminating between fast and slow states is small. In this case, 
\begin{equation}
K_0^{(C)} - K_0^{(B)} = K_0^{(B)} \frac{\Delta \epsilon}{k_B T_0} + \mathcal{O}(\epsilon) \qquad \qquad k_0^{(C)} - k_0^{(B)} = k_0^{(B)} \frac{\Delta \epsilon}{k_B T_0} + \mathcal{O}(\epsilon)
\end{equation}
Using this expansion we can compute the higher moments of the transition rates, and also the expressions for the quantities defined in Eq.~\eqref{vars} up to the first order in $\Delta \epsilon$. Then, Eq.~\eqref{2order}, up to the first order in $\Delta \epsilon$, becomes:
\begin{equation}
P^{(2)}_C - P^{(2)}_B = \frac{\Delta E}{T_0^4} P^{eq}_B L_s^2 \left( 1 - 2 L_s \tanh \left( \frac{1}{2 L_s} \right) \right) \Delta \epsilon > 0
\end{equation}
with $L_s = \sqrt{D/(k_{B \to A} + 2 k_{A \to B})}$. As expected this difference vanishes when $D \to 0$, while, in the opposite limit $D \to \infty$, the characteristic length scale $L_s$ tends to a constant value.

\subsection*{Scaling parameter $\sqrt{D/k_{A \to B}}$}

Here we highlight the role played by the scaling parameter $\sqrt{D/k_{A \to B}}$, which is similar to what naturally arises as a characteristic length scale for the system, $L_s$, from the second order solution.

Although we have seen that the state $C$ is globally favourable, for small values of the gradient $\nabla T$, it is possible to find a region of parameter space in which:
\begin{equation}
P(C(x)) - P(B(x)) < 0
\label{c1c2}
\end{equation}
For sake of simplicity, let us assume that the forward reactions $k_{A \to B}$ is much faster than the reverse one $k_{B \to A}$, i.e. the energy difference $\Delta E$ is large. In this case, it is easy to see that, up to the first order in $\nabla T$, we get:
\begin{equation}
p^{(1)}_C(x) - p^{(1)}_B(x) = (M^+_B - M^+_C) \left( 2 L^- \frac{\sinh\left( \frac{1-2x}{2 L^-} \right)}{\cosh\left( \frac{1}{2 L^-} \right)} \right)
\end{equation}
This quantity descreases along the temperature gradient, meaning that it is at its lowest value and negative for $x = 1$. However, its integrated value still remains positive as we have shown above. In particular, the ratio between the second order and the first order contributions is:
\begin{equation}
\frac{p^{(2)}_C(x)-p^{(2)}_B(x)}{p^{(1)}_C(x) - p^{(1)}_B(x)} \propto \frac{1}{T_0} \sqrt{\frac{D}{k_{A \to B}}}
\end{equation}
involving the characteristic length scale $L_s$ introduced above, when $\Delta E$ is large.

\subsection*{Numerical analysis beyond perturbation theory}

In the main text we have proposed an argument to show that the unbalance between $C$ and $B$ at stationarity presents a maximum for a given $\Delta T^*$. Here we corroborate this motivation with numerical simulations.

We find the stationary solution of the Chapman-Kolmogorov equation \eqref{cha-kol} numerically with no-flux boundary conditions, using the built-in solver of Mathematica. The ``accuracy goal" has been set equal to half the Machine Precision (53 bits). We consider the presence of a linear temperature gradient, $D_A = D_B = D_C = D$, $\Delta E = E_A - E_B$, and $E_B = E_C$, which are the working conditions of the manuscript. In Fig. 1A we fix $T(0) = 1$, $\Delta E = 2$, and show the ratio $P_C/P_B$ as a function of $\Delta T$ for four different combinations of $\Delta \epsilon$ and the diffusion coefficient $D$.

\begin{figure}[h]
\centering
\includegraphics[width=1 \columnwidth]{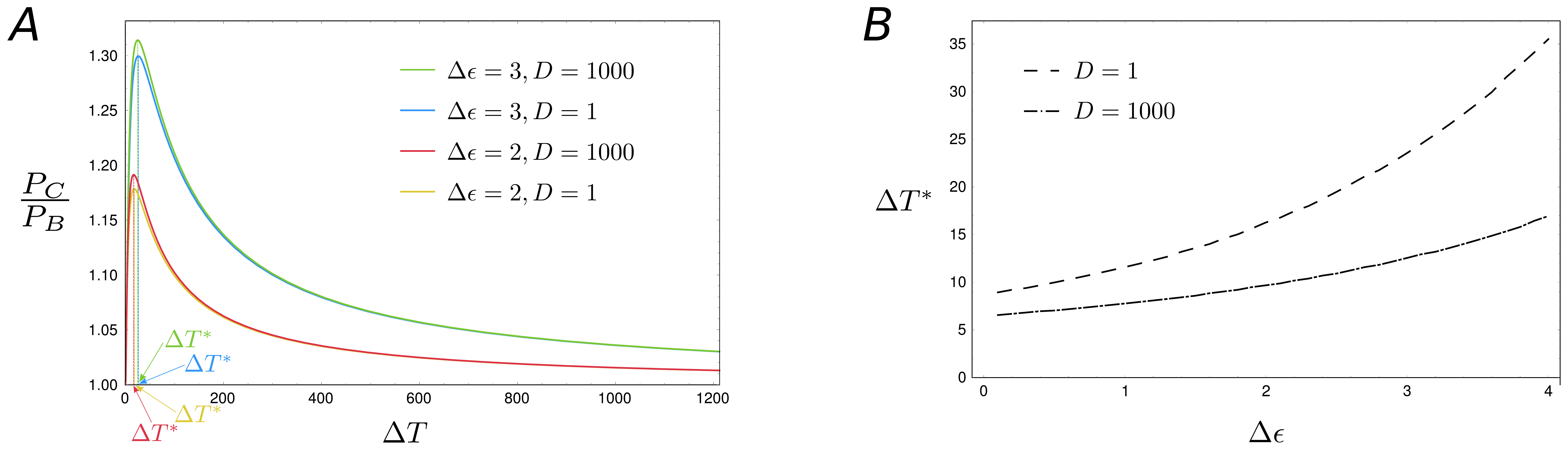}
\caption{A) Unbalance between $P_C$ and $P_B$ in the continuous model, quantified through the ratio $P_C/P_B$, as a function of the gradient $\Delta T$. We set $T(0) = 1$, $\Delta E = 2$ and two different values for the energy barrier between $A$ and $C$, $\Delta \epsilon$, and the diffusion coefficient $D$. In particular, $\Delta \epsilon = 3$, $D = 1000$ (green curve), $\Delta \epsilon = 3$, $D = 1$ (blue curve), $\Delta \epsilon = 2$, $D = 1000$ (red curve), and $\Delta \epsilon = 2$, $D = 1$ (yellow curve). The unbalance reaches a maximum for a finite value of $\Delta T = \Delta T^*$, and then goes back to $1$ asymptotically for infinite gradients. B) $\Delta T^*$ as a function of $\Delta \epsilon$ for $D = 1$ (dashed curve) and $D = 1000$ (dot-dashed curve), $T(0) = 1$ and $\Delta E = 2$. When the barrier increases, the optimal value of the gradient icreases as well. Moreover, the diffusion favours the overcome of energy barriers in non-equilibrium conditions.}
\end{figure}

\begin{figure}[bh]
\centering
\includegraphics[width=0.65 \columnwidth]{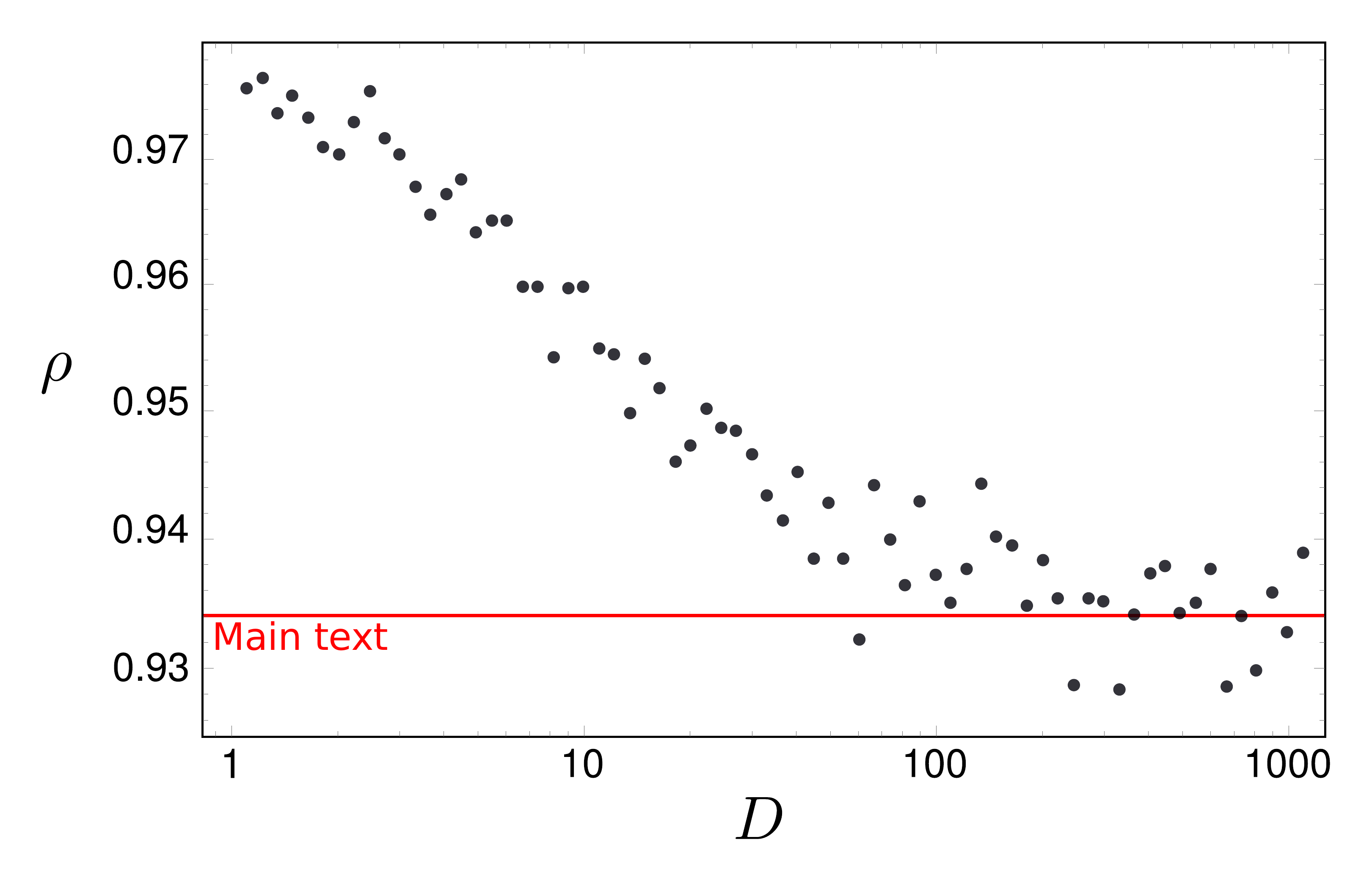}
\caption{Correlation coefficient $\rho$ between $R_{CB}$ and $\dot{S}/DE$ as a function of the diffusion coefficient $D$ (the $x$-axis is in log-scale). The parameters have been set as in the main text of manuscript: $T_2 = 1$, $\Delta \epsilon = 3$ and $10^3$ values of $\Delta E$ and $\Delta T$, drawn from a Gaussian distribution with unitary mean and standard deviation $0.1$. The red line indicates $\rho$ for the particular realization shown in Fig. 2 of the main text in the case of infinite diffusion. Black points indicates the correlation for one realization of $\Delta E$ and $\Delta T$ with a given and finite $D$.}
\end{figure}

As discussed in the main text, the ratio between the populations of $C$ and $B$ starts from $1$ for $\Delta T = 0$ (equilibrium conditions), increases up to a maximum value, and then goes back to $1$ asymptotically for infinite gradients.

\subsubsection*{``Optimal" gradient}

The ratio $R_{CB} = P_C/P_B$ reaches its maximum value, \textit{i.e.} the out-of-equilibrium selection is maximized, for a given value of the gradient, $\Delta T^*$, that we name ``optimal". 

In Fig. 1B we study numerically how the optimal gradient varies as a function of the energy barrier $\Delta \epsilon$, which is responsible for the kinetic symmetry-breaking of the system. As expected, as the value of barrier increases, $\Delta T^*$ increases as well, since more thermal energy is required to overcome the barrier and let the system relaxes to the fastest state $C$ out of equilibrium. However, if $\Delta T > \Delta T^*$ the system starts becoming more and more insensitive to the energy barriers, relaxing into an equilirium state in the limit of infinite gradient.

Moreover, we show two curves of $\Delta T^*(\Delta \epsilon)$, for two different values of the diffusion coefficient $D$. We note that, for a given value of the available energy, $\Delta T$, a stronger diffusion reduces the value of $\Delta T^*$, thus allowing the particles to overcome more easily energy barriers and favouring the non-equilibrium selection of fastest species.


\section*{\MakeUppercase{Relation between selection and dissipation}}

This section is dedicated to a more in-depth numerical study of the correlation between the unbalance of the populations of species $C$ and $B$, $R_{CB}$, and the steady state entropy production, $\dot{S}/\Delta E$ \cite{schnakenberg, bamos2}, for the two box model presented in the manuscript.

In Fig. 2 of main text we have shown the correlation between the unbalance of the population of $C$ with respect to the one of $B$, quantified through their ratio $R_{CB}$, and the entropy production at stationarity, fixing a specific value for each parameter, most notably for the energy barrier $\Delta \epsilon$. Moreover, in the main text we presented only the case of infinite diffusion. Here we discuss the robustness of our results for other choices of the paramters.

\begin{figure}[h]
\centering
\includegraphics[width=1 \columnwidth]{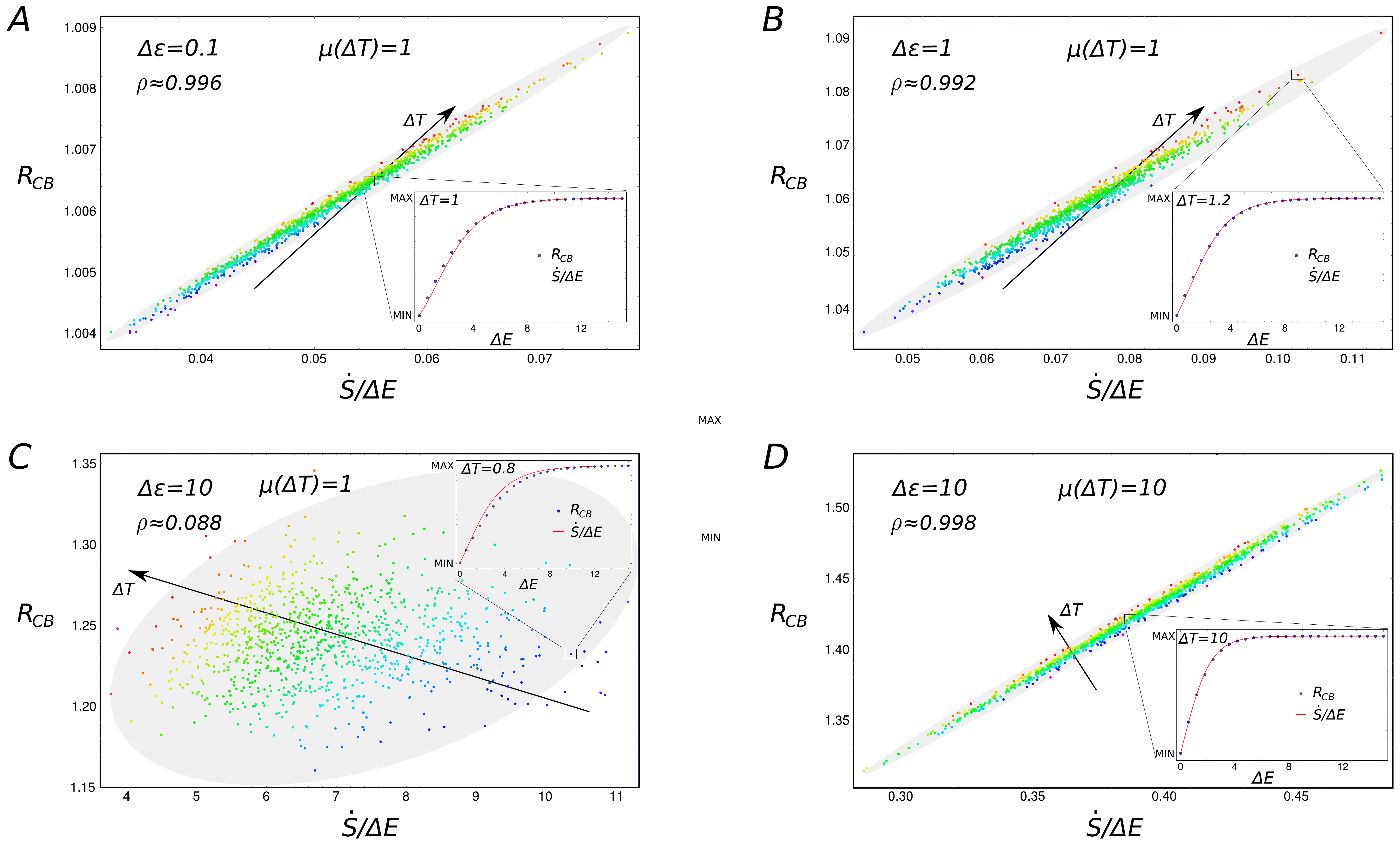}
\caption{Correlation between $R_{CB} = P_C/P_B$ and $\dot{S}/\Delta E$ for infinite diffusion, $T_2 = 1$, a set of $10^3$ values of $\Delta E$ drawn from a Gaussian distribution with mean $1$ and standard deviation $0.1$, a set of $10^3$ values of $\Delta T$ Gaussian distributed with mean $\mu(\Delta T)$ and standard deviation $0.1$, for four different choices for the energy barrier $\Delta \epsilon$. $\rho$ is the correlation coefficient. A) $\Delta \epsilon = 0.1$ and $\mu(\Delta T) = 1$. The correlation is close to unity. In the Inset the behaviour of $R_{CB}$ and $\dot{S}/DE$ (rescaled to lie in the same range) is presented for a fixed value $\Delta T = 1$, and a wide range of energies $\Delta E$. B) $\Delta \epsilon = 1$ and $\mu(\Delta T) = 1$, exhibiting $\rho$ close to $1$. The Inset present the correlation in the particular case $\Delta T = 1.2$. C) $\Delta \epsilon = 10$ and $\mu(\Delta T) = 1$. The correlation is lost for higher values of energy barriers in the whole range of $\Delta T$, which becomes particularly effective in determining the value of $R_{CB}$. In fact, the correlation is still present for fixed values of the gradient. In the Inset the case $\Delta T = 0.8$ is shown. D) The correlation is recovered for high energy barriers, when the mean of the distribution of the gradients is raised ($\mu(\Delta T) = 10$). The Inset shows the correlation for $\Delta T$ fixed to $10$ between $R_{CB}$ and $\dot{S}/\Delta E$ rescaled to lie in the same range.}
\end{figure}

First of all, in Fig. 2, we show that the correlation is not significatively affected by the presence of finite diffusion, evidencing the fact that we can study only the limit $d \to +\infty$ to obtain useful insights on the system.

The energy barrier $\Delta \epsilon$ determines the kinetics of the system, which becomes relevant at stationarity in non-equilibrium conditions (as extensively discussed in the main text), playing a fundamenatal role in our model. In Fig. 3 we show the correlation between $R_{CB}$ and $\dot{S}/\Delta E$ for three different choices of $\Delta \epsilon$. When energy barriers are equal or less than the value of $T_2$, the correlation coefficient is close to unity (Fig.s 3A and 3B). Conversely, when $\Delta \epsilon \gg k_B T_2$, the correlation breaks down when the whole explored range of gradients is considered (Fig. 3C). Notably, fixing a specific value of $\Delta T$, the correlation is recovered, as shown in Fig. 3D. This evidence can be interpreted as follows: when energy barriers are greater than the thermal energy of the system, the value of the gradient becomes crucial to determine what fraction of particles eventually overcome them. This is reflected by the features of Fig. 3C: similar colors, \textit{i.e.} similar values of $\Delta T$, are arranged in parallel lines, providing a signature for the presence of a correlation with $R_{CB}$. In other words, for peculiar settings of the system (i.e. high energy barrier, Fig. 3C), the correlation is robust for a smaller range of gradient with respect to other situations (i.e. low energy barrier, Fig.s 3A and 3B, or strong thermal gradient, Fig. 3D).

\section*{SELECTION IN A BRANCHING TREE OF REACTIONS}

\begin{figure}[b]
\centering
\includegraphics[width=0.6 \columnwidth]{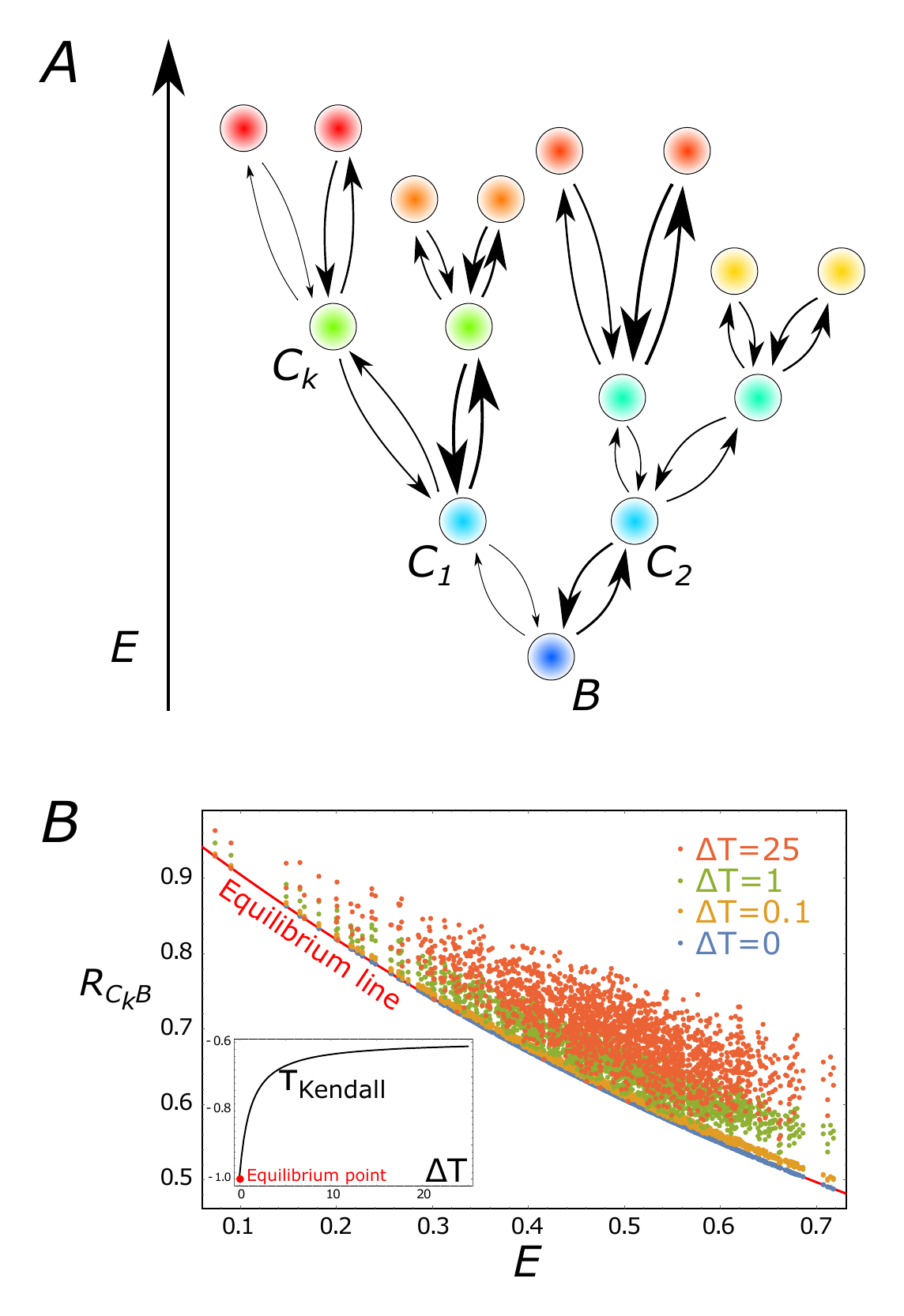}
\caption{A) A chain of reactions branching from lower to higher energies with a branching ratio equal to $2$. Same colors correspond to the same energy, and the thicker is the arrow, the faster the reaction. B) Population of states with respect to the lowest energy state $B$, as a function of the energy $E$, for a system with $10$ branching levels ($2047$ states). The solid red line indicates the equilibrium case $\Delta T = 0$. Points with different colors refer to different values of $\Delta T$. In the \textit{Lower Inset} the Kendall correlation coefficient \cite{kendall} between $R_{C_kB}$ and $E_k$ is shown. We set $k_B = 1, T=1$, $E_B = 0$, and, for each three-state subsystem, $\Delta E \sim U(0,0.1)$ and $\Delta \epsilon \sim U(10,50)$. As the system is driven away from equilibrium, the reaction rates, and not only the energies, become important to determine the steady population.}
\label{fig:4}
\end{figure}

In the main text we have inspected how global localization phenomena amy arise in a chain of connected reactions in presence of fast diffusion. Another interesting and quite simple example in which the fast reactions play a leading role is provided by a branching tree of chemical reactions from lower to higher energies, as sketched in Fig.~\ref{fig:4}A. Also in this case we consider the infinitely fast diffusion limit. When no temperature gradient is applied, the population of each state follows the Boltzmann distribution, and progressively deviates from it as the temperature gradient is increased, the role of the reaction rates becoming progressively more important (see Fig.~\ref{fig:4}B). 
As an indicator of this feature, we compute the Kendall correlation coefficient \cite{kendall} between $R_{C_k B}$ and $E_k$ for each state. As expected, it is equal to $-1$ at equilibrium, whereas it lowers when increasing $\Delta T$. This means that population and energy tend to become less correlated in a non-equilibrium stationary state. Moreover, the effect is enhanced when the system operates far from equilibrium, as shown in the Inset of Fig.~\ref{fig:4}B.

Again, the velocity of each reaction with respect to its adjacent ones is not sufficient to determine the stationary population of the states involved in it. In fact, the most important ingredient determining which states $\{C_k^*\}$ will have a net increase in their populations, in the non-equilibrium steady state, is once again the weight of the path connecting each $C_k^*$ to a reference state ($B$ in this case; clearly, as can be shown, the ranking of the states does not depend on the chosen reference). Stated otherwise, the system select the fastest paths from lower ($B$) to higher energies ($C_k$), ideally providing a natural identification of the best possible tree of reactions which lead to the most stable metastable states.

\section*{\MakeUppercase{Thermophoresis in a two-species two-box model}}

In the main text, we have discussed the emergence of thermophoresis \cite{duhr, piazza}, \textit{i.e.} the accumulation of particles to the hot or cold side of a gradient, as an inevitable consequence of the law of thermodynamics. In particular, we have shown that this phenomenon may appear in a three-state two-box model, where the fastest state ($C$) is selected in non-equilibrium steady state, introducing an unbalance between the transport coefficients of each species. We also provided an interpretation of thermophoresis as a selection process \textit{in real space}, rather than in the state of the species.

\begin{figure}[h]
\centering
\includegraphics[width=1\columnwidth]{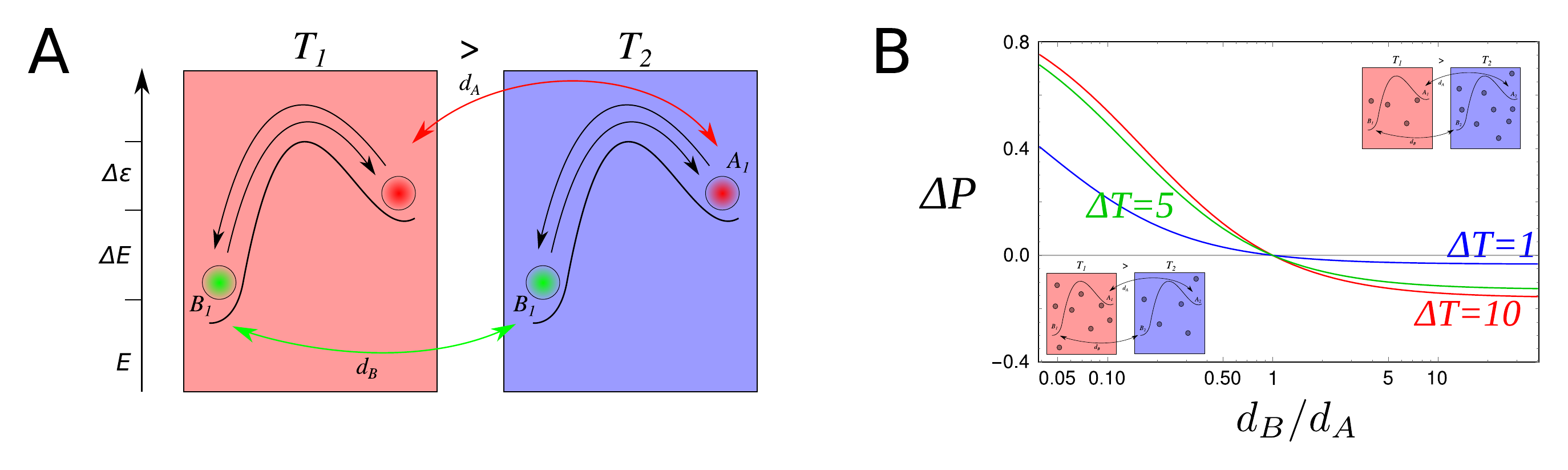}
\caption{A) A two-state chemical system diffusing in a temperature gradient, modeled as two connected boxes at different temperatures, $T_1 > T_2$. The transport coefficient of each species is equal to $d_X$, where $X=A,B$. B) Unbalance of populations $\Delta P = P_2 - P_1$ as a function of $\alpha = d_B/d_A$ (in logscale) for three different values of the gradient. The system can accumulate both on the warm or cold side depending on the ratio between the transport rates. The higher is the available energy in the form of a thermal gradient, the greater will be the unbalance. The parameters have been set as follows: $T = 1, \Delta E = 5, d_A = 1$, and $k_B = 1$ for sake of simplicity.}
\label{fig:4}
\end{figure}

Here, we want to point out that thermphoresis is independent of the selection of the fastest species, being intimately related to the kinetic-symmetry breaking of the transport properties instead. To this aim, we consider a two species two-box model, as shown in Fig. 5A. Here, by construction, the selection of states is prevented: there is trivially only one dissipative cycle in the system passing through all the states.

However, if $d_A \neq d_B$, an unbalance between $P_1 = P(A_1) + P(B_1)$ and $P_2 = P(A_2) + P(B_2)$ is obtained (see Fig. 5B), which is the emerging behaviour reminiscent of thermophoresis we are looking for. Indeed, to this unbalance we can associate a Soret coefficient \cite{platten, soret1, soret2}, following the same strategy explained in the main text. In the limit of small diffusion, we have the following simple expression as a function of energy difference $\Delta E$, temperature $T_2$ and ratio of transport rates $d_B/d_A$:
\begin{equation}
S_T = \frac{(1-d_B/d_A) e^{- \Delta E/k_B T_2}}{(1 + e^{- \Delta E/k_B T_2}) ((d_B/d_A) e^{- \Delta E/k_B T_2}  + 1) k_B T_2^2}
\end{equation}


\end{widetext}

\end{document}